\documentclass[11pt]{elsarticle}

\pdfoutput=1

\makeatletter
\def\ps@pprintTitle{%
  \let\@oddhead\@empty
  \let\@evenhead\@empty
  \let\@oddfoot\@empty
  \let\@evenfoot\@oddfoot
}
\makeatother

\usepackage{url}
\usepackage{breakurl}
\usepackage[breaklinks,
            colorlinks = true,
            linkcolor = blue,
            urlcolor  = blue,
            citecolor = blue,
            anchorcolor = blue]{hyperref}

\usepackage{graphicx}
\usepackage[margin=1.25in]{geometry}
\usepackage[usenames,dvipsnames]{color}

\newcommand\acp{\begin{center}
\rule[-0.2in]{\hsize}{0.01in}\\\rule{\hsize}{0.01in}\\
\vskip 0.1in Submitted to the  Proceedings\\ 
of the African Conference on Fundamental and Applied Physics
    \vskip 0.05in
    {\it Second Edition, ACP2021, March 7--11, 2022 --- Virtual Event}\\
\rule{\hsize}{0.01in}\\\rule[+0.2in]{\hsize}{0.01in} \\
\end{center}}

\usepackage[firstpage=true]{background}
\backgroundsetup{contents={\parbox{6.5in}{\acp}}, scale=1,placement=top,opacity=1,color=black,position={3.25in,1.2in}}

\usepackage{fancyhdr}
\fancypagestyle{plain}{%
  \fancyhf{}%
  \fancyhead[C]{}
  \fancyfoot[C]{\thepage}
}

\fancypagestyle{empty}{%
  \fancyhf{}%
  \fancyhead[C]{{\it ACP2021, March 7--11, 2022 --- Virtual Edition}}
  \fancyfoot[C]{\thepage}
}
\usepackage [english]{babel}
\usepackage [autostyle, english = american]{csquotes}
\MakeOuterQuote{"}

\begin{document}

\begin{frontmatter}


\title{Physics Masterclasses in Africa and the World}

\author[add1]{Uta Bilow}
\ead{Uta.Bilow@tu-dresden.de}
\author[add2]{Kenneth Cecire}
\ead{kcecire@nd.edu}

\address[add1]{Institut für Kern- und Teilchenphysik, Technische Universität Dresden, Dresden, 01062, Germany}
\address[add2]{Department of Physics and Astronomy, University of Notre Dame, Notre Dame, 46556, Indiana, USA}

\begin{abstract}
\noindent 
International Masterclasses (IMC) enable high school students and teachers to work with particle physicists to analyze authentic data from contemporary experiments and experience being “physicists for a day”. The IMC program has a worldwide reach, including several universities and research institutes in Egypt, Algeria, Morocco, Sao Tomé and Principe, and South Africa. As technical infrastructure in Africa improves, there is a great opportunity for many more African institutes to offer IMC on their premises. The authors will discuss the advantages of IMC to Africa, how institutes may join, and ways to overcome obstacles.
\end{abstract}


\end{frontmatter}

%



\section{Introduction}
\label{sec:intro}
\noindent
International Masterclasses (IMC) are experimental particle physics education and outreach events for high school students aged 15 to 19 years. In the IMC concept, students become "scientists for a day". They are invited to a research institute or a university, where they attend introductory talks on topics such as the Standard Model, detectors, and accelerators to help them prepare them to make their own measurements. Students analyze authentic data from an actual experiment to be able to answer a research question or to draw conclusions from the measurement. Currently, students at different masterclasses are engaged in measurements from the Large Hadron Collider (LHC) at Conseil Européen pour la Recherche Nucléaire (CERN), Belle II at Kō Enerugī Kasokuki Kenkyū (KEK), neutrino experiments at Fermilab, and particle therapy from Gesellschaft für Schwerionenforschung (GSI). The capstone of the masterclass is an international videoconference with 3-5 other masterclass institutes that have made measurements from the same experiment.  ~\cite{imcsite}~\cite{art}

What makes this event a \textit{masterclass} is the interaction of high school students and their teachers with the particle physicists who host and tutor the event. Most important, as students analyze data for their measurements, they interact with physicists to ask questions and clarify the meaning of the data. It is in this interaction that the physicists impart not only specific information but also their approach to understanding both the experiments and the results. One way to express the goal of a masterclass is to help students see data as scientists see data.

This paper is organized as follows. In Sections 2, we explain details of the  masterclass day. In Section 3, we discuss goals, motivations, and requirements. In Section 4, we flesh out the opportunities from different "flavors" of masterclass. We conclude in Section 5 with an invitation to participate. 

\section{The Moving Parts}
\label{sec:parts}
\noindent 
The masterclass day varies depending on the institute and time zone but generally follows a plan similar to this Sample Agenda:
\begin{center}
\begin{tabular}{||c c||} 
 \hline
 \textbf{Local Time} & \textbf{Activity} \\ [0.5ex] 
 \hline\hline
 (prep) & (classroom) \\ 
 \hline
 09:00-09:30 & registration and welcome \\
 \hline
 09:30-10:30 & introduction to particle physics \\
 \hline
 10:30-11:30 & second talk or tour \\
 \hline
 11:30-12:00 & introduction to measurement \\ 
 \hline
 12:00-13:00 & lunch (with physicists) \\
 \hline
 13:00-15:00 & data analysis \\
 \hline
 15:00-16:00 & local combination of results and discussion \\
 \hline
 16:00-17:00 & international videoconference \\
 \hline
\end{tabular}
\end{center}
\noindent
The international videoconference has the following key features:
\begin{itemize}
    \item 45-60 minute duration
    \item 3-5 institutes, reflecting international collaboration (where possible)
    \item Same measurement, different data
    \item 2-3 moderators (physicists, graduate students)
    \item moderation centers: CERN, Fermilab, KEK, GSI, TriIniversity Meson Facility (TRIUMF)
    \item Agenda of videoconference:
    \begin{itemize}
        \item welcome
        \item combination and discussion of results
        \item general Q\&A
        \item quiz (CERN).
    \end{itemize}
\end{itemize}
~\cite{imcsite}~\cite{pm2022}
\section{Why and What}
\label{sec:2w}
\noindent
Why do we facilitate and promote International Masterclasses? Why do physicists invest their time and efforts in them? IMC is a way to help start development of the next generation of scientists and engineers by giving young learners a real experience with cutting-edge science. In the same way, IMC promotes interest in science among all students. Visiting a university or laboratory where they work side-by-side with physicists near their homes as well as international collaborators, high school students have the chance to participate in the excitement of contemporary physics research. Teachers who bring students to masterclasses year after year often develop their own expertise and interest that can enhance their instruction in the classroom.

To host a masterclass, local organizers need to secure space to work with students and computers with high-speed internet (unless the masterclass is local only). All masterclass packages are web browser-based (although some need installation of Java and software, e.g. event display). All masterclass packages are available for educational purposes without cost. The physicist leadership of the masterclass should consist of an organizer, at least one tutor for each ten students, coordination with IMC, some flexibility to deal with local conditions and unforeseen problems, and a welcoming attitude.
\section{Opportunities}
\label{opp}
\noindent
The masterclasses described so far have been from the IMC "regular season". These are the mostly standard full masterclasses that occur each year from February to April and are coordinated by IMC central. However, there are additional opportunities that arise each year.

World Wide Data Day (W2D2) is a one-day simplified masterclass event held in November or December each year. W2D2 measurements are from the ATLAS and CMS detectors in the LHC and can be completed in about two hours, from introduction to videoconference end. High school teachers organize W2D2 measurements at their own schools and connect their students to physicists in the videoconferences, which occur in a day-long "shift".~\cite{w2d2}

IMC also holds special masterclasses for the International Day of Women and Girls in Science on 11th February each year. These are much like standard masterclasses but are designed to especially benefit young women with female scientists as role models and discussions of women in physics.~\cite{imcsite}

In addition, any organizers can host masterclasses anytime of the year, in or out of the IMC schedule. IMC gives these initiatives support upon request and many institutions, from universities and laboratories to high schools, have successfully created meaningful particle physics masterclasses opportunities for students.

IMC is open to alternative designs of masterclasses as needed by local organizers. There have been masterclasses for university physics students, masterclasses as outreach activities in conferences, and many more examples. 
\section{Conclusion and Invitation}
\label{ci}
\noindent
International Masterclasses have spread to countries all around the world. Africa is part of IMC but there is great room for mutual growth and enrichment for high school students and teachers between IMC and physicists in Africa. Some of these opportunities include masterclasses already described, adaptations to make the masterclasses more appropriate in the local context, and masterclasses redesigned for African physics experiments, such as HESS or the African Light Source.

The authors invite physicists and graduate students in Africa and the world to think about organizing a masterclass, tutoring in a masterclass, becoming a videoconference moderator, or even creating a new masterclass measurement. The authors, as co-Coordinators of International Masterclasses, are eager to help.

\bibliographystyle{elsarticle-num}
\bibliography{myreferences} 

\end{document}